\documentclass[preprintnumbers,amsmath,amssymbm,prl]{revtex4}
\usepackage{epsfig}
\usepackage{graphicx}

\begin{document}
\title{Black-hole quasinormal resonances: Wave analysis versus a geometric-optics approximation}
\author{Shahar Hod}
\address{The Ruppin Academic Center, Emeq Hefer 40250, Israel}
\address{ }
\address{The Hadassah Institute, Jerusalem 91010, Israel}
\date{\today}

\begin{abstract}
\ \ \ It has long been known that null unstable geodesics are
related to the characteristic modes of black holes-- the so called
quasinormal resonances. The basic idea is to interpret the free
oscillations of a black hole in the eikonal limit in terms of null
particles trapped at the unstable circular orbit and slowly leaking
out. The real part of the complex quasinormal resonances is related
to the angular velocity at the unstable null geodesic. The imaginary
part of the resonances is related to the instability timescale (or
the inverse Lyapunov exponent) of the orbit. While this
geometric-optics description of the black-hole quasinormal
resonances in terms of perturbed null {\it rays} is very appealing
and intuitive, it is still highly important to verify the validity
of this approach by directly analyzing the Teukolsky wave equation
which governs the dynamics of perturbation {\it waves} in the
black-hole spacetime. This is the main goal of the present paper. We
first use the geometric-optics technique of perturbing a bundle of
unstable null rays to calculate the resonances of near-extremal Kerr
black holes in the eikonal approximation. We then directly solve the
Teukolsky wave equation (supplemented by the appropriate physical
boundary conditions) and show that the resultant quasinormal
spectrum obtained directly from the {\it wave} analysis is in accord
with the spectrum obtained from the geometric-optics approximation
of perturbed null {\it rays}.
\end{abstract}
\bigskip
\maketitle


The no-hair conjecture \cite{Wheeler} asserts that the external
field of a perturbed black hole relaxes to a Kerr-Newman spacetime,
characterized solely by three parameters: the black-hole mass,
charge, and angular momentum. This implies that perturbation fields
left outside the black hole would either be radiated away to
infinity, or be swallowed by the black hole.

The relaxation phase in the dynamics of perturbed black holes is
characterized by `quasinormal ringing', damped oscillations with a
discrete spectrum (see e.g. \cite{Nollert1} for reviews). At late
times, all perturbations are radiated away in a manner reminiscent
of the last pure dying tones of a ringing bell
\cite{Press,Cruz,Vish,Davis}. Quasinormal resonances are expected to
play a prominent role in gravitational radiation emitted by a
variety of astrophysical scenarios involving black holes. Being the
characteristic `sound' of the black hole itself, these free
oscillations are of great importance from the astrophysical point of
view. They allow a direct way of identifying the spacetime
parameters, especially the mass and angular momentum of the black
hole.

The dynamics of black-hole perturbations is governed by the
Regge-Wheeler equation \cite{RegWheel} in the case of a spherically
symmetric Schwarzschild black hole, and by the Teukolsky equation
\cite{Teu} for rotating Kerr-Newman spacetimes. The black hole
quasinormal modes (QNMs) correspond to solutions of the wave
equations with the physical boundary conditions of purely outgoing
waves at spatial infinity and purely ingoing waves crossing the
event horizon \cite{Detwe}. Such boundary conditions single out a
discrete set of black-hole resonances $\{\omega_n\}$ (assuming a
time dependence of the form $e^{-i\omega t}$). In analogy with
standard scattering theory, the QNMs can be regarded as the
scattering resonances of the black-hole spacetime. They thus
correspond to poles of the transmission and reflection amplitudes of
a standard scattering problem in a black-hole spacetime.

In accord with the spirit of the no-hair conjecture \cite{Wheeler},
the external perturbation fields would either fall into the black
hole or radiate to infinity. This implies that the perturbation
decays with time and the corresponding QNM frequencies are therefore
{\it complex}. It turns out that there exist an infinite number of
quasinormal modes, characterizing oscillations with decreasing
relaxation times (increasing imaginary part), see
\cite{Leaver,Noll2,Hodln3,KeshHod} and references therein. The mode
with the smallest imaginary part (known as the fundamental mode)
determines the characteristic dynamical timescale for generic
perturbations to decay \cite{GlaAnd2,Hod1,Hod2,Gruz,Hodb,Hod8}.

In most cases of physical interest \cite{Noterot}, the black-hole
QNMs must be computed {\it numerically} by solving the Teukolsky
equation supplemented by the appropriate physical boundary
conditions. However, Mashhoon \cite{Mash} has suggested an {\it
analytical} technique of calculating the QNMs in the
geometric-optics (eikonal) limit. The basic idea is to interpret the
black-hole free oscillations in terms of null particles trapped at
the unstable circular orbit and slowly leaking out
\cite{CarMir,Mash,Goeb}. The real part of the complex quasinormal
resonances is related to the angular velocity at the unstable null
geodesic while the imaginary part of the resonances is related to
the instability timescale of the orbit (or the inverse Lyapunov
exponent of the geodesic \cite{CarMir}).

It should be emphasized that, Mashhoon's approach of computing
black-hole quasinormal frequencies in the eikonal limit is somewhat
indirect: instead of explicitly solving the Teukolsky wave equation
which describes the dynamics of perturbation {\it waves} in the
black-hole spacetime, Mashhoon analyzed perturbations of test null
{\it rays} in the unstable circular orbit of the black hole. While
this geometric-optics description of black-hole quasinormal
resonances in terms of perturbed null {\it rays} is very appealing
and intuitive, it is still important to verify the validity of this
approach by directly solving the Teukolsky master equation which
governs the dynamics of perturbation {\it waves} in the Kerr
black-hole spacetime. Filling this gap is the main goal of the
present paper. We shall first use the geometric-optics technique of
perturbing a bundle of unstable null geodesics to calculate the
resonances of near-extremal Kerr black holes in the eikonal limit.
We shall then directly solve the Teukolsky wave equation
(supplemented by the appropriate physical boundary conditions) and
show that the resultant resonances of the wave equation agree with
those obtained from Mashhoon's indirect approach of perturbing null
rays in the appropriate eikonal approximation.

We start with Mashhoon's approach of calculating the black-hole QNM
resonances in the eikonal limit $l=m \gg 1$, where $l$ is the
angular-momentum of the perturbed null rays, and $m$ is the
azimuthal harmonic index of the rays. According to Mashhoon's
analysis of perturbed null geodesics, the Kerr QNM frequencies in
the $l=m \gg 1$ limit are given by \cite{Mash} (we use natural units
in which $G=c=\hbar=1$)
\begin{equation}\label{Eq1}
\omega_n =m\omega_+  -i(n+{1\over 2})\beta\omega_+\ \ \ ; \ \ \
n=0,1,2,...\ ,
\end{equation}
where
\begin{equation}\label{Eq2}
\omega_+\equiv {{M^{1/2} \over{r^{3/2}_{ph}+aM^{1/2}}}}\ ,
\end{equation}
is the Kepler frequency for null rays in the unstable equatorial
circular orbit of the black hole, and
\begin{equation}\label{Eq3}
r_{ph}\equiv 2M\{1+\cos[{2 \over 3}\cos^{-1}(-a/M)]\}\  ,
\end{equation}
is the limiting circular photon orbit. The function $\beta$ is given
by
\begin{equation}\label{Eq4}
\beta= {{(12M)^{1/2}(r_{ph}-r_+)(r_{ph}-r_-)} \over
{r^{3/2}_{ph}(r_{ph}-M)}}\  ,
\end{equation}
see Ref. \cite{Mash} for details. Here $M$ and $a$ are the mass and
angular momentum per unit mass of the black hole, respectively.

We now focus on the near extremal limit $T_{BH} \to 0$, where
\begin{equation}\label{Eq5}
T_{BH}={{(M^2-a^2)^{1/2}}\over{4\pi M[M+(M^2-a^2)^{1/2}]}}\  ,
\end{equation}
is the Bekenstein-Hawking temperature of the black hole. Let
$r_{\pm}=M\pm\epsilon$ with $\epsilon/M \ll 1$, where
$r_{\pm}=M+(M^2-a^2)^{1/2}$ are the black-hole event and inner
horizons. The Bekenstein-Hawking temperature of the black hole now
reads $T_{BH}={\epsilon/{4\pi M^2}}+O(\epsilon^2/M^3)$. After some
algebra we find from Eq. (\ref{Eq3})

\begin{equation}\label{Eq6}
r_{ph}=M+{2\epsilon \over {\sqrt{3}}} +O(\epsilon^2/M)\  .
\end{equation}
This, in turn, implies
\begin{equation}\label{Eq7}
\omega_+ ={1 \over {2M}}-{{\sqrt{3} \epsilon}\over
{4M^2}}+O(\epsilon^2/M^3)\ ,
\end{equation}
and
\begin{equation}\label{Eq8}
\beta={\epsilon \over M} +O(\epsilon^2/M^2).
\end{equation}
Substituting Eqs. (\ref{Eq7}) and (\ref{Eq8}) into Eq. (\ref{Eq1}),
one finds that the black-hole quasinormal resonances in the eikonal
approximation $l=m>>1$ are given by
\begin{equation}\label{Eq9}
\omega_n =m\Omega+2\pi T_{BH}(1-{{\sqrt 3}\over 2})m- i2\pi
T_{BH}(n+{1\over 2})+O(MT^2_{BH})\ ,
\end{equation}
where
\begin{equation}\label{Eq10}
\Omega \equiv {a \over {r^2_+ +a^2}}={1\over{2M}}-2\pi
T_{BH}+O(MT^2_{BH})\ ,
\end{equation}
is the angular velocity of the black-hole event horizon.

We shall now study directly the Teukolsky wave equation in order to
determine the fundamental (least-damped) resonant frequencies of the
Kerr black hole. The Teukolsky equation is amenable to an {\it
analytical} treatment in the near-extremal limit $(M^2-a^2)^{1/2}\ll
a\lesssim M$.

In order to determine the black-hole resonances we shall analyze the
scattering of massless waves in the Kerr spacetime. The dynamics of
a perturbation field $\Psi$ in the rotating Kerr spacetime is
governed by the Teukolsky equation \cite{Teu}. One may decompose the
field as
\begin{equation}\label{Eq11}
\Psi_{slm}(t,r,\theta,\phi)=e^{im\phi}S_{slm}(\theta;a\omega)\psi_{slm}(r)e^{-i\omega
t}\ ,
\end{equation}
where $(t,r,\theta,\phi)$ are the Boyer-Lindquist coordinates,
$\omega$ is the (conserved) frequency of the mode, $l$ is the
spheroidal harmonic index, and $m$ is the azimuthal harmonic index
with $-l\leq m\leq l$. The parameter $s$ is called the spin weight
of the field, and is given by $s=\pm 2$ for gravitational
perturbations, $s=\pm 1$ for electromagnetic perturbations, $s=\pm
{1\over 2}$ for massless neutrino perturbations, and $s=0$ for
scalar perturbations. (We shall henceforth omit the indices $s,l,m$
for brevity.) With the decomposition (\ref{Eq11}), $\psi$ and $S$
obey radial and angular equations, both of confluent Heun type
\cite{Heun,Flam}, coupled by a separation constant $A(a\omega)$.

The angular functions $S(\theta;a\omega)$ are the spin-weighted
spheroidal harmonics which are solutions of the angular equation
\cite{Teu,Flam}
\begin{equation}\label{Eq12}
{1\over {\sin\theta}}{\partial \over
{\partial\theta}}\Big(\sin\theta {{\partial
S}\over{\partial\theta}}\Big)+\Big[a^2\omega^2\cos^2\theta-2a\omega
s\cos\theta-{{(m+s\cos\theta)^2}\over{\sin^2\theta}}+s+A\Big]S=0\ .
\end{equation}
The angular functions are required to be regular at the poles
$\theta=0$ and $\theta=\pi$. These boundary conditions pick out a
discrete set of eigenvalues $A_l$ labeled by an integer $l$. [In the
$a\omega\ll 1$ limit these angular functions become the familiar
spin-weighted spherical harmonics with the corresponding angular
eigenvalues $A=l(l+1)-s(s+1)+O(a\omega)$.] The angular equation
(\ref{Eq12}) can be solved analytically in the $l=m>>1$ limit to
yield
\begin{equation}\label{Eq13}
A=m^2+O(m)\  ,
\end{equation}
see Ref. \cite{Marolf}.

The radial Teukolsky equation is given by
\begin{equation}\label{Eq14}
\Delta^{-s}{{d}
\over{dr}}\Big(\Delta^{s+1}{{d\psi}\over{dr}}\Big)+\Big[{{K^2-2is(r-M)K}\over{\Delta}}
-a^2\omega^2+2ma\omega-A+4is\omega
r\Big]\psi=0\ ,
\end{equation}
where $\Delta\equiv r^2-2Mr+a^2$ and $K\equiv (r^2+a^2)\omega-am$.
For the scattering problem one should impose physical boundary
conditions of purely ingoing waves at the black-hole horizon and a
mixture of both ingoing and outgoing waves at infinity (these
correspond to incident and scattered waves, respectively). That is,
\begin{equation}\label{Eq15}
\psi \sim
\begin{cases}
e^{-i\omega y}+{\mathcal{R}}(\omega)e^{i \omega y} & \text{ as }
r\rightarrow\infty\ \ (y\rightarrow \infty)\ ; \\
{\mathcal{T}}(\omega)e^{-i (\omega-m\Omega)y} & \text{ as }
r\rightarrow r_+\ \ (y\rightarrow -\infty)\ ,
\end{cases}
\end{equation}
where the ``tortoise" radial coordinate $y$ is defined by
$dy=[(r^2+a^2)/\Delta]dr$. The coefficients ${\cal T}(\omega)$ and
${\cal R}(\omega)$ are the transmission and reflection amplitudes
for a wave incident from infinity. They satisfy the usual
probability conservation equation $|{\cal T}(\omega)|^2+|{\cal
R}(\omega)|^2=1$.

The discrete quasinormal frequencies are the scattering resonances
of the black-hole spacetime. They thus correspond to poles of the
transmission and reflection amplitudes. (The pole structure reflects
the fact that the QNMs correspond to purely outgoing waves at
spatial infinity.) These resonances determine the ringdown response
of a black hole to external perturbations. Teukolsky and Press
\cite{TeuPre} and also Starobinsky and Churilov \cite{StaChu} have
analyzed the black-hole scattering problem in the double limit $a\to
M$ and $\omega\to m\Omega$. Detweiler \cite{Det} then used that
solution to obtain a resonance condition for near-extremal Kerr
black holes. Define
\begin{equation}\label{Eq16}
\sigma\equiv {{r_+-r_-}\over{r_+}}\ \ ;\ \ \tau\equiv
M(\omega-m\Omega)\ \ ;\ \ \hat\omega\equiv \omega r_+\  .
\end{equation}
Then the resonance condition obtained in \cite{Det} for $\sigma<<1$
and $\tau<<1$ is:
\begin{equation}\label{Eq17}
-{{\Gamma(2i\delta)\Gamma(1+2i\delta)\Gamma(1/2+s-2i\hat\omega-i\delta)\Gamma(1/2-s-2i\hat\omega-i\delta)}\over
{\Gamma(-2i\delta)\Gamma(1-2i\delta)\Gamma(1/2+s-2i\hat\omega+i\delta)\Gamma(1/2-s-2i\hat\omega+i\delta)}}=
(-2i\hat\omega\sigma)^{2i\delta}
{{\Gamma(1/2+2i\hat\omega+i\delta-4i\tau/\sigma)}\over{\Gamma(1/2+2i\hat\omega-i\delta-4i\tau/\sigma)}}\
,
\end{equation}
where $\delta^2\equiv 4\hat\omega^2-1/4-A-a^2\omega^2+2ma\omega$.
Taking cognizance of Eq. (\ref{Eq13}), one finds
\begin{equation}\label{Eq18}
\delta={{\sqrt{3}\over 2}}m+O(1)\  ,
\end{equation}
for near-extremal Kerr black holes in the $l=m>>1$ limit.

The left-hand-side of Eq. (\ref{Eq17}) has a well defined limit as
$a\to M$ and $\omega\to m\Omega$. We denote that limit by ${\cal
L}$. Since $\delta\simeq \sqrt{3}m/2>>1$, one has
$(-i)^{-2i\delta}=e^{-i\sqrt{3}m\ln(-i)}=e^{-i\sqrt{3}m\ln
e^{-i\pi/2}}=e^{-i\sqrt{3}m(-i\pi/2)}=e^{-\sqrt{3}\pi m/2}\ll 1$,
which implies $\lambda\equiv(-2i\hat\omega\sigma)^{-2i\delta}\ll 1$.
Thus, a consistent solution of the resonance condition, Eq.
(\ref{Eq17}), may be obtained if
$1/\Gamma(1/2+2i\hat\omega-i\delta-4i\tau/\sigma)=O(\lambda)$.
Suppose
\begin{equation}\label{Eq19}
1/2+2i\hat\omega-i\delta-4i\tau/\sigma=-n+\eta\lambda+O(\lambda^2)\
,
\end{equation}
where $n\geq 0$ is a non-negative integer and $\eta$ is an unknown
constant to be determined below. Then one has
\begin{equation}\label{Eq20}
\Gamma(1/2+2i\hat\omega-i\delta-4i\tau/\sigma)\simeq\Gamma(-n+\eta\lambda)\simeq
(-n)^{-1}\Gamma(-n+1+\eta\epsilon)\simeq\cdots\simeq [(-1)^n
n!]^{-1}\Gamma(\eta\lambda)\  ,
\end{equation}
where we have used the relation $\Gamma(z+1)=z\Gamma(z)$
\cite{Abram}. Next, using the series expansion
$1/\Gamma(z)=\sum_{k=1}^{\infty} c_k z^k$ with $c_1=1$ [see Eq.
$(6.1.34)$ of \cite{Abram}], one obtains
\begin{equation}\label{Eq21}
1/\Gamma(1/2+2i\hat\omega-i\delta-4i\tau/\sigma)=(-1)^n
n!\eta\lambda+O(\lambda^2)\  .
\end{equation}
Substituting this into Eq. (\ref{Eq17}) one finds $\eta={\cal
L}/[(-1)^n n!\Gamma(-n+2i\delta)]$.

Finally, substituting $4\tau/\sigma=(\omega-m\Omega)/2\pi T_{BH}$,
$2i\hat\omega=im+O(mMT_{BH})$ for $\omega=m\Omega+O(mT_{BH})$, and
$\delta=\sqrt{3}m/2+O(1)$ into Eq. (\ref{Eq19}), one obtains the
resonance condition
\begin{equation}\label{Eq22}
(\omega-m\Omega)/2\pi T_{BH}=i[-n+\eta\lambda-1/2+im({{\sqrt{3}\over
2}}-1)]\ .
\end{equation}
The black-hole quasinormal resonances in the $l=m>>1$ limit are
therefore given by the formula
\begin{equation}\label{Eq23}
\omega_n=m\Omega+2\pi T_{BH}(1-{{\sqrt{3}\over 2}})m-i2\pi
T_{BH}(n+{1\over 2})+O(MT^2_{BH})\ ,
\end{equation}
where $n=0,1,2,...$ . We emphasize that this result, obtained from
the direct wave analysis, coincides with the previously derived
spectrum Eq. (\ref{Eq9}) of the ray analysis (the geometric-optics
approximation).

In summary, we have studied analytically the quasinormal spectrum of
rapidly-rotating Kerr black holes. We first used the technique of
perturbing a bundle of unstable equatorial null geodesics to
calculate the quasinormal resonances of Kerr black holes in the
eikonal limit $l=m>>1$. We then used the alternative (and the more
direct) approach of solving the Teukolsky wave equation which
governs the dynamics of perturbation waves in the Kerr black-hole
spacetime. We have shown that the resonance spectrum (\ref{Eq23})
obtained directly from the {\it wave} analysis is in accord with the
spectrum (\ref{Eq9}) which was obtained from the geometric-optics
approximation of perturbed null {\it rays}.

\bigskip
\noindent
{\bf ACKNOWLEDGMENTS}
\bigskip

This research is supported by the Meltzer Science Foundation. I
thank Uri Keshet, Yael Oren, Liran Shimshi and Clovis Hopman for
helpful discussions. I also thank Andrei Gruzinov for interesting
correspondence. It is my pleasure to thank Steve Detweiler for
valuable comments.



\begin{thebibliography}{99}
\bibitem{Noterot} It is well known that realistic stellar objects generally rotate
about their axis, and are therefore not spherical. Thus, an
astrophysically realistic model of wave dynamics in black-hole
spacetimes must involve a non-spherical background geometry with
angular momentum.
\end{thebibliography}
\end{document}